\documentclass[twocolumn,showpacs,preprintnumbers,amsmath,amssymb]{revtex4}

\usepackage{mathrsfs}
\usepackage{amsmath}
\usepackage{amssymb}
\usepackage[dvipdfm]{graphicx}
\usepackage{bm}

\usepackage{dcolumn}
\usepackage{bm}

\begin{document}

\preprint{KUNS-2029}

\title{Hawking Radiation from Rotating Black Holes and Gravitational Anomalies}

\author{Keiju Murata}
\email{murata@tap.scphys.kyoto-u.ac.jp}
\author{Jiro Soda}%
 \email{jiro@tap.scphys.kyoto-u.ac.jp}
\affiliation{
 Department of Physics,  Kyoto University, Kyoto 606-8501, Japan
}%

\date{\today}


\begin{abstract}
 We study the Hawking radiation from Rotating black holes 
  from gravitational anomalies point of view. First, we
 show that the scalar field theory near the Kerr black hole horizon can 
 be reduced to the 2-dimensional effective theory. 
 Then, following Robinson and Wilczek,
 we derive the Hawking flux by requiring the cancellation of
 gravitational anomalies.
 We also apply this method to Hawking radiation from higher dimensional
 Myers-Perry black holes. In the Appendix, we present
  the trace anomaly derivation of Hawking radiation  
 to argue the validity of the boundary condition at the horizon.
\end{abstract}

\maketitle

\section{Introduction}

Understanding of physics of the black hole horizon, 
such as the black hole entropy and the Hawking radiation, 
is believed to be a hint for the quantum theory of gravity. 
Many efforts have been devoted to this theme. Recently, there was a progress
in understanding of black hole entropy  
\cite{Carlip:2006fm}. 
There, the breakdown of the diffeomorphism symmetry at the horizon, 
namely anomalies,  played an important role.   
Since the Hawking radiation \cite{Hawking:1974sw} as well as the black hole entropy 
is the property inherent in the horizon,
 it is natural to expect that 
 the Hawking radiation is also associated with
 anomalies. 

Many years ago, Christiansen and
Fulling found that Hawking radiation can be derived from the trace anomaly
\cite{Christensen:1977jc} in the case of 
(1+1)-dimensional Schwarzschild spacetime. 
In this approach, as is usual, boundary conditions both
at the horizon and at the infinity are required to specify the vacuum. 
Hence, it is difficult to attribute 
the Hawking radiation to the property of the event horizon.
It should be also mentioned that the method is not applicable to more than
(2+1)-dimension.
Recently,  Robinson and Wilczek suggested
a new derivation of Hawking radiation from Schwarzschild black holes through  
gravitational anomalies \cite{Robinson:2005pd}. It should be noted that this
derivation is applicable to any dimension.
In their work, the Hawking radiation is understood
as compensating flux to cancel gravitational anomalies at the
horizon. The advantage of this derivation is that it requires
 the information only at the horizon. 
 In order to prove that this gravitational anomaly method is relevant,
 we need to show the universality of it. 
  Concerning this, Iso et al. have shown 
 that the Hawking radiation from Reissner-Nordstrom black
holes can be explained as the flux which cancel 
 gravitational and $U(1)$ gauge anomalies~\cite{Iso:2006wa}. 
 They have also clarified the boundary condition at the horizon.

In this paper, we further extend  Robinson and Wilczek's
derivation of Hawking radiation \cite{Robinson:2005pd}  to
rotating black holes. We also discuss the boundary condition at the horizon 
by comparing the gravitational anomaly method with
 the trace anomaly method~\cite{Christensen:1977jc}. 

The organization of this paper is as follows:
in sec.II, we review the gravitational anomaly method.
In sec.III, we apply the gravitational anomaly method to Kerr black holes.
In sec.IV, higher dimensional black holes, the so-called Meyers-Perry solutions,
are considered. The final section is devoted to the conclusion.
In the Appendix, we present the trace anomaly method to argue the
validity of the boundary condition at the horizon.

\section{\label{sec:level1}Hawking radiation and gravitational
 anomalies}
In this section, we will review the gravitational anomaly 
method~\cite{Robinson:2005pd,Iso:2006wa} to make the paper self-contained. 

Consider the metric of the type,
\begin{equation}
 ds^2 = -f(r)dt^2 + \frac{1}{f(r)}dr^2 + r^2 d\Omega_{D-2}^2 \ ,
\end{equation}
where $f(r)$ is some function which admits the event horizon.
The horizon is located at $r=r_H$, where $f(r_H)=0$. The surface
gravity is given by $\kappa=\frac{1}{2}\partial_r f |_{r_H}$. 
First, we show that
the scalar field theory on this metric can be reduced to the 2-dimensional
theory. The action of the scalar field is
\begin{equation}
\begin{split}
 S[\varphi] =& \frac{1}{2}\int d^D x \sqrt{-g}\,\varphi \nabla^2 \varphi\\
=& \frac{1}{2}\int d^D x\,r^{D-2} \sqrt{\gamma} \\
 &\times \varphi \left(
 -\frac{1}{f}\partial_t^2 + \frac{1}{r^{D-2}}\partial_r r^{D-2} f
 \partial_r + \frac{1}{r^2} 
 \Delta_{\Omega}\right) \varphi  \ ,
\end{split}
\end{equation}
where $\gamma$ is the determinant of $d\Omega_{D-2}^2$ 
and $\Delta_\Omega$ is the
collection of the angular derivatives. Now we take the limit  
$r \rightarrow r_H$ and leave only dominant terms. 
Thus, the action becomes
\begin{equation}
 \begin{split}
 S[\varphi] =& \frac{{r_H}^{D-2}}{2}\int d^D x \sqrt{\gamma} \,\varphi \left(
 -\frac{1}{f}\partial_t^2 + \partial_r f \partial_r
  \right) \varphi\\
=& \sum_n \frac{{r_H}^{D-2}}{2} \int dt dr \,\varphi_n \left(
 -\frac{1}{f}\partial_t^2 + \partial_r f \partial_r
  \right) \varphi_n   
\end{split}
\end{equation}
in the second line $\varphi$ is expanded by $(D-2)$-dimensional
spherical harmonics. This action is infinite set of the scalar fields on
the 2-dimensional metric
\begin{equation}
 ds^2 = -f(r)dt^2 + \frac{1}{f(r)}dr^2   \ .
\end{equation}
Thus, we can reduce the scalar field theory in $D$-dimensional black hole 
spacetime to that in 2-dimensional spacetime near the horizon.

In this 2-dimensional spacetime, we treat the black hole horizon as the
boundary of the spacetime and discard ingoing modes 
near the horizon because these ingoing
modes cannot affect the dynamics of the scalar fields out of horizon. 
This 2-dimensional
theory is chiral. Let us split the region into two ones: 
$r_H \leq r \leq r_H + \epsilon$ where the theory is chiral and 
$r_H + \epsilon \leq r$ where the theory is not chiral. We
will take the limit $\epsilon \rightarrow 0$ ultimately. It is known
that the gravitational anomaly arises in 2-dimensional chiral theory and
its explicit form takes \cite{Kimura:1970iv,Alvarez-Gaume:1983ig,Bertlmann:2000da}
\begin{equation}
 \nabla_\mu {T^\mu}_{\nu} = -\frac{1}{96\pi \sqrt{-g}}\epsilon^{\beta
  \delta}\partial_\delta \partial_\alpha \Gamma^\alpha_{\nu\beta} \ ,
\label{eq:anom}
\end{equation} 
where the convention $\epsilon^{01}=+1$ is used. 
We define $A_\nu$ and ${N^\mu}_\nu$ as
\begin{equation}
 \nabla_\mu {T^\mu}_{\nu} \equiv A_\nu \equiv
  \frac{1}{\sqrt{-g}}\partial_\mu {N^\mu}_\nu   \ .
\label{eq:anom2}
\end{equation}
In the region of $r_H + \epsilon \leq r$, we have $A_\nu = {N^\mu}_\nu = 0$. 
But in the near horizon, 
$r_H \leq r \leq r_H + \epsilon$, the components of these are
\begin{equation}
\begin{split}
 {N^t}_t &= {N^r}_r = 0   \ , \\
 {N^r}_t &= -\frac{1}{192\pi}({f'}^2+f''f)   \ , \\
 {N^t}_r &= -\frac{1}{192\pi f^2}(f'^2-f''f)  \ , 
\end{split}
\end{equation}
and
\begin{equation}
 \begin{split}
  A_t &= -\frac{1}{192\pi}(f'^2 + f''f)'    \ , \\
  A_r &= 0  \ ,
 \end{split}
\label{eq:anom3}
\end{equation}
where $' \equiv \partial_r$ denotes the derivative with respect to $r$.

The effective action for the metric $g_{\mu\nu}$ after
integrating out the scalar field is
\begin{equation}
 W[g_{\mu\nu}]=-i \ln \left(\int \mathscr{D} \phi\,e^{i S[\phi,\,g_{\mu\nu}]}\right)
\end{equation}
where $S[\phi,\,g_{\mu\nu}]$ is the classical action. The 
infinitesimal general coordinate transformation 
\begin{eqnarray}
  x^\mu \longrightarrow x^\mu - \lambda^\mu \ 
\end{eqnarray}
induces the variation of the effective action
\begin{equation}
\begin{split}
 -\delta_\lambda W =& \int d^2x \sqrt{-g} \,\lambda^\nu \nabla_\mu
  \{{{T_{(H)}}^\mu}_\nu H(r) + {{T_{(o)}}^\mu}_\nu \Theta_+(r)\}\\
 =& \int d^2x \,\lambda^t \{\partial_r({N^r}_t H)\\
&\qquad + ({{T_{(o)}}^r}_t -
 {{T_{(H)}}^r}_t + {N^r}_t)\delta(r-r_H -\epsilon)\}\\ 
 &+ \int d^2 x \,\lambda^r ({{T_{(o)}}^r}_r -
 {{T_{(H)}}^r}_r)\delta(r-r_H -\epsilon)  \ ,
\end{split}
\label{eq:GCT}
\end{equation}
where $\Theta_+ (r) = \Theta(r - r_H -\epsilon)$ and 
$H(r) = 1 -\Theta_+(r)$. The subscript $H$ and $o$ represent the value in the
region $r_H \leq r \leq r_H + \epsilon$ and $r_H + \epsilon \leq r$, 
respectively. 
To obtain the last result, we used the fact that ${{T_{(o)}}^\mu}_\nu$
 is covariantly conserved and ${{T_{(H)}}^\mu}_\nu$
obeys anomalous equation
(\ref{eq:anom}). 

 Taking into account the time
independence of ${{T_{(H)}}^\mu}_\nu$ and ${{T_{(o)}}^\mu}_\nu$,
we can integrate Eq. (\ref{eq:anom2}) as
\begin{equation}
 \begin{split}
  {T^t}_t &= -\frac{K+Q}{f} - \frac{B(r)}{f} - \frac{I(r)}{f} 
               + {T^\alpha}_\alpha(r)  \ , \\
  {T^r}_r &= \frac{K+Q}{f} + \frac{B(r)}{f} + \frac{I(r)}{f} \ , \\
  {T^r}_t &= - K + C(r) = -f^2{T^t}_r \ ,
 \end{split}
\label{eq:EM}
\end{equation}
where $T$ represents $T_{(H)}$ or $T_{(o)}$, 
\begin{equation}
 \begin{split}
  C(r) &\equiv \int^r_{r_H}A_t(r')dr'  \ ,\\
  B(r) &\equiv \int^r_{r_H}f(r')A_r(r')dr'  \ ,\\
  I(r) &\equiv \frac{1}{2}\int^r_{r_H}{T^\alpha}_\alpha(r')f'(r')dr'  \ ,
 \end{split}
\end{equation}
 and $K$
and $Q$ are constants of integration. We denote the constants by $K_H,Q_H$ and
$K_o,Q_o$ at the region $r_H \leq r \leq r_H + \epsilon$ and 
$r_H + \epsilon \leq r$ respectively.
From Eq. (\ref{eq:anom3}), $B(r)$ should be zero.
In the limit $r \rightarrow r_H$, we have  $C(r)\rightarrow 0$ and 
$I(r)/f \rightarrow\frac{1}{2}{T^\alpha}_\alpha(r_H)$. 
Thus,  Eq. (\ref{eq:EM}) becomes
\begin{equation}
 \begin{split}
  {T^t}_t &= -\frac{K + Q}{f} + \frac{1}{2}{T^\alpha}_\alpha(r) \ ,\\
  {T^r}_r &= \frac{K + Q}{f} + \frac{1}{2}{T^\alpha}_\alpha(r) \ ,\\
  {T^r}_t &= - K  = -f^2{T^t}_r \ .
 \end{split}
\label{eq:EM2}
\end{equation}
The four constants $K_H$, $K_o$, $Q_H$ and $Q_o$ have to be
determined to cancel the gravitational anomaly.
Substituting Eq. (\ref{eq:EM2}) into
Eq. (\ref{eq:GCT}) and taking the limit 
$\epsilon \rightarrow 0$, we obtain
\begin{equation}
\begin{split}
-\delta_\lambda W =& \int d^2x \,\lambda^t \{\partial_r({N^r}_t H)\\
&\qquad + (-K_H + K_o + {N^r}_t)\delta(r-r_H)\}\\ 
 &+ \int d^2 x \,\lambda^r \frac{K_H + Q_H - K_o - Q_o}{f}\delta(r-r_H) \ .
\end{split}
\end{equation}
In order to keep the diffeomorphism invariance,
this variation should  vanish. Since the first term cannot be canceled by
delta function terms, it should be canceled by the quantum effect
of the ingoing modes. Therefore, we ignore the first term.
Setting $\delta_\lambda W = 0$, we get
\begin{equation}
 \begin{split}
  K_o &=  K_H - \Phi \ ,\\
  Q_o &= Q_H + \Phi \ ,
 \end{split}
\label{eq:constsrel}
\end{equation}
where
\begin{equation}
 \Phi \equiv \left. \frac{{f'}^2}{192\pi} \right|_{r=r_H} 
            = \frac{\kappa^2}{48\pi}   \ .
\label{eq:flux}
\end{equation}
We need to know $K_H$ to obtain the Hawking flux. For this purpose, we
adopt the boundary condition proposed in \cite{Iso:2006wa}. 
Let us introduce the covariant
energy momentum tensor $\tilde{T}_{\mu\nu}$ which satisfies
the covariant anomaly equation
\begin{equation}
\nabla_\mu {\tilde{T}^\mu}_{\nu} = \frac{1}{96\pi
 \sqrt{-g}}\epsilon_{\mu\nu} \partial^\mu R \ .
\label{eq:covanom}
\end{equation}
 We impose the boundary condition on the covariant anomalous energy
 momentum tensor ${\tilde{T}_{(H)}}^{\mu\nu}$, 
 since the boundary condition at the horizon should be diffeomorphism 
invariant.  The boundary condition we take is 
\begin{equation}
 \tilde{T}^{\quad r}_{(H)\ t} 
    = {{T_{(H)}}^r}_t - \frac{1}{192\pi}(ff''-2{f'}^2) = 0 \ .
\end{equation}
We discuss the validity of this boundary condition 
in the Appendix. 
Thus, we obtain $K_H = 2\Phi$ and therefore
\begin{equation}
 {{T_{(o)}}^r}_t = -\Phi  \ .
\end{equation}
 So, $\Phi$ is the flux of Hawking radiation.
The flux of black body radiation in 2-dimension is 
$\Phi = \frac{\pi}{12}T^2$. Comparing this with (\ref{eq:flux}),
we get the correct Hawking temperature of the black hole
\begin{equation}
 T = \frac{\kappa}{2\pi} \ .
\end{equation}
It is important to extend the above analysis 
to more realistic rotating black holes.

\section{\label{sec:level2}Hawking radiation from Kerr black holes}
In this section, we will show that the Hawking radiation from Kerr
black holes can be understood as the flux cancelling the gravitational anomaly.
The point is that, near the horizon, the scalar field theory in 4-dimensional
Kerr black hole spacetime can be reduced to the 2-dimensional field theory. 
As the spacetime is not spherically symmetric, this is an unexpected result.

In Boyer-Linquist coordinates, Kerr metric reads
\begin{multline}
 ds^2 = -\frac{\Delta-a^2 \sin^2 \theta}{\Sigma}dt^2 - 2a\sin^2 \theta
  \frac{r^2+a^2-\Delta}{\Sigma}dtd\phi\\
  + \frac{(r^2+a^2)^2-\Delta a^2
  \sin^2 \theta}{\Sigma}\sin^2 \theta d\phi^2 +
  \frac{\Sigma}{\Delta}dr^2 + \Sigma d\theta^2 
\end{multline}
where 
\begin{equation}
 \begin{split}
  \Sigma &= r^2 + a^2 \cos^2 \theta \ , \\
  \Delta &= r^2 - 2Mr + a^2 \\
         &= (r-r_+)(r-r_-)\ .
 \end{split}
\end{equation}
The outer and inner horizon are located at $r=r_+,r_-$ 
respectively.
The determinant of the metric is
\begin{equation}
 \sqrt{-g} = \Sigma \sin \theta  \ ,
\end{equation}
and the inverse of the metric of ($t,\phi$) parts is
\begin{equation}
 \begin{split}
  g^{tt} &= -\frac{(r^2+a^2)^2-\Delta a^2 \sin^2\theta}{\Sigma\Delta}  \ ,\\
  g^{\phi\phi} &=
  \frac{\Delta-a^2\sin^2\theta}{\Sigma\Delta\sin^2\theta}  \ , \\
  g^{t\phi} &= -\frac{a(r^2+a^2-\Delta)}{\Sigma\Delta}  \ .
 \end{split}
\end{equation}
The action for the scalar field in the Kerr spacetime is
\begin{equation}
\begin{split}
 S[\varphi] =& \frac{1}{2}\int d^4 x \sqrt{-g}\,\varphi \nabla^2 \varphi \\
=& \frac{1}{2}\int d^4 x\sqrt{-g} \,\varphi\frac{1}{\Sigma}
 \left[-\left(\frac{(r^2+a^2)^2}{\Delta}-a^2 \sin^2 \theta
 \right)\partial_t^2 \right.\\
 &- \frac{2a(r^2 + a^2 - \Delta)}{\Delta}\partial_t
 \partial_\phi + \left(\frac{1}{\sin^2
 \theta}-\frac{a^2}{\Delta}\right)\partial_\phi^2\\
 &+ \left. \partial_r \Delta
 \partial_r + \frac{1}{\sin\theta}\partial_\theta \sin \theta
 \partial_\theta \right]\varphi  \ .
\end{split}
\end{equation}
Taking the limit $r \rightarrow r_+$ and leaving the dominant terms, we have
\begin{equation}
\begin{split}
 S[\varphi] =& \frac{1}{2}\int d^4x\sin\theta \,\varphi
 \left[ -\frac{(r_+^2+a^2)^2}{\Delta}\partial_t^2 \right.\\
 &- \left.\frac{2a(r_+^2 +
 a^2)}{\Delta}\partial_t \partial_\phi
 -\frac{a^2}{\Delta}\partial_\phi^2
 + \partial_r \Delta \partial_r
 \right]\varphi  \ .
\label{eq:dalembertian}
\end{split}
\end{equation}
Now we transform the coordinates to the locally non-rotating coordinate
system by
\begin{equation}
\begin{cases}
 \psi = \phi - \Omega_H t \ , \\
 \xi = t  \ ,
\end{cases}
\end{equation}
where
\begin{equation}
 \Omega_H \equiv \frac{a}{r_+^2 + a^2}  \ .
\end{equation}
 Using $(\xi,r,\theta,\psi)$ coordinates,
 we can rewrite the action (\ref{eq:dalembertian}) as
\begin{equation}
 S[\varphi] = \frac{a}{2\Omega_H}\int d^4x \sin\theta \,\varphi \left( -\frac{1}{f(r)}\partial_\xi^2 + \partial_r f(r) \partial_r \right)\varphi \ ,
\end{equation}
where 
\begin{equation}
 f(r) \equiv \frac{\Omega_H \Delta}{a} \ .
\end{equation}
One can see the angular derivative terms disappear completely.
The spherical harmonics expansion 
$\varphi(x) = \sum_{l,m} \varphi_{l\,m}(\xi,r)Y_{l\,m}(\theta,\psi)$
finally gives the effective 2-dimensional action
\begin{equation}
\begin{split}
 S[\varphi] =& \frac{a}{\Omega_H}\sum_{l,m}\frac{1}{2}\int d\xi dr\\
&\times \varphi_{l\,m}
  \left(-\frac{1}{f(r)}\partial_\xi^2 + \partial_r f(r) \partial_r
  \right) \varphi_{l\,m}  \ .
\end{split}
\end{equation}
The effective 2-dimensional metric can be read off from the above action
as 
\begin{equation}
 ds^2 = -f(r)d\xi^2 + \frac{1}{f(r)}dr^2 \ .
\end{equation}
Thus, we have reduced the 4-dimensional field theory to the 2-dimensional one.
This 2-dimensional metric tells us that, near the horizon, the
geometry of Kerr spacetime is the Rindler spacetime when $r_+ > r_-$.
 In the extremal case $r_+ = r_-$, the near horizon geometry reduces
 to AdS$_2$ which is consistent with the result of \cite{Bardeen:1999px}. 

Now we can derive the Hawking radiation from Kerr black holes
 using the formalism explained in Sec. \ref{sec:level1}.
 The flux determined by the anomaly cancellation arguments is given by
 Eq.(18) from which we can read off the temperature as 
\begin{equation}
 \begin{split}
  T&=\frac{1}{4\pi} \partial_r f|_{r_+}\\
   &= \frac{r_+^2-a^2}{4\pi r_+(r_+^2 + a^2)} = \frac{\sqrt{M^2-a^2}}{4\pi M(M+\sqrt{M^2-a^2})} \ .
 \end{split}
\end{equation}
This is nothing but the Hawking temperature of Kerr black holes. 
Although we have not yet shown that
this flux is  Plankian, we assume this is so. 
In the view of 4-dimensional theory, the distribution function is
$(\exp(\omega/T)-1)^{-1}$ near
the horizon in the ($\xi,\psi$) coordinates. In this coordinate, the scalar
field with energy 
$\omega$ and axial quantum number $m$ is 
$\varphi \propto \exp(i\omega \xi + im\psi)$. In the ($t,\phi$) coordinates, 
$\varphi \propto \exp(i(\omega-m\Omega_H)t + im\phi))$. Hence, in
($t,\phi$) coordinates,we obtain the distribution function 
\begin{equation}
 \frac{1}{\exp((\omega-m\Omega_H)/T)-1} \ .
\end{equation}
Thus, we get the correct chemical potential for Kerr black holes.
It is also easy to calculate the flux of the angular momentum
from this result. 

In the case of Kerr-Newman black holes with the electric charge $Q$ and
the magnetic charge $P$, the Hawking temperature of the neutral scalar field can
be calculated similarly by replacing $a^2 \rightarrow a^2+Q^2+P^2$.

\section{the case of Myers-Perry black holes}
The discussion in the previous section 
can be extended to Myers-Perry black holes with
only one rotating axis. The Myers-Perry metric in $D$-dimension is
\cite{Myers:1986un,Frolov:2006ib}
\begin{equation}
\begin{split}
 ds^2 =& -dt^2+\frac{Udr^2}{V-2M}+\frac{2M}{U}(dt+\sum^n_{i=1}a_i \mu_i^2
  d\phi_i)^2\\
&+\sum^n_{i=1} (r^2+a_i^2)(\mu_i^2 d\phi_i^2 + d\mu_i^2)+\epsilon r^2
 d\mu_{n+\epsilon}^2 
\end{split}
\end{equation}
where
\begin{equation}
 \begin{split}
  V &= r^{\epsilon-2}\prod^n_{i=1}(r^2+a_i^2)  \ , \\
  U &= V(1-\sum^n_{i=1}\frac{a_i^2 \mu_i^2}{r^2+a_i^2}) \ ,
 \end{split}
\end{equation}
and $n$ is the integer part of
$(D-1)/2$ and $\epsilon = 1(D\text{:even}),\,0 (D\text{:odd})$. The
coordinates $\mu_i$ are not independent but obey the relation
\begin{equation}
 \sum^n_{i=1} \mu_i^2 + \epsilon \mu_{n+\epsilon}^2=1 \ .
\end{equation}
We consider the Myers-Perry black hole with one rotating axis. Denote
$a_1 = a,\, a_i =0\,(\text{for}\,i \neq 1),\,\mu_1 = \mu$ and $\phi_1 = \phi$. 
Using the polar coordinates
\begin{equation}
 \begin{split}
  \mu =& \cos \theta \ ,\\
  \mu_2 =& \sin \theta \cos \theta_2  \ , \\
  \mu_3 =& \sin \theta \sin \theta_2 \cos \theta_3 \ , \\
&\vdots  
 \end{split}
\end{equation}
we can write the metric as
\begin{equation}
\begin{split}
 ds^2 =& \left(-1+\frac{2M}{U}\right)dt^2 + \mu^2
 \left(r^2+a^2+\frac{2Ma^2\mu^2}{U} \right)
 d\phi^2\\
&+ \frac{4Ma \mu^2}{U} dtd\phi + \frac{U dr^2}{V-2M}\\
&+ \frac{(r^2+a^2)U}{V}d\theta^2 + r^2 d\gamma^2 \ ,
\end{split}
\end{equation}
where
\begin{equation}
 d\gamma^2 \equiv \sin^2 \theta d\Omega_{n+\epsilon-2}^2
 + (\mu_2^2d\phi_2^2 + \dots + \mu_n^2 d\phi_n^2) \ ,
\end{equation}
and $d\Omega_{n+\epsilon-2}^2$ is the metric of $S^{n+\epsilon-2}$ with 
coordinates ($\theta_2,\cdots,\theta_{n+\epsilon-1}$).
The inverse of the metric of ($t,\phi$) parts is given by
\begin{equation}
\begin{split}
g^{tt}&= -\frac{V\{(r^2+a^2)U+2a^2\mu^2 M\}}{(r^2+a^2)(V-2M)U} \ , \\
g^{\phi\phi}&=\frac{(U-2M)V}{\mu^2(r^2+a^2)(V-2M)U} \ , \\
g^{t\phi}&= \frac{2aMV}{(r^2+a^2)(V-2M)U}  \ ,
\end{split}
\end{equation}
and the determinant of the metric is
\begin{equation}
 \sqrt{-g}=\frac{\mu(r^2+a^2)U}{V}r^{D-4}\sqrt{\gamma} \ .
\end{equation}
Note that the horizon is located at $r=r_+$, determined by $V(r=r_+) = 2M$.
The ($t,r,\phi$) parts and 
($\theta,\theta_2,\dots,\theta_{n+\epsilon-1},\phi_2,\dots,\phi_n$)
parts of the metric are decoupled and the 
inverse metric of ($\theta,\theta_2,\cdots,\theta_{n+\epsilon-1},\phi_2,\cdots,\phi_n$) parts are
nonsingular at the horizon, so these are negligible in the scalar field
action near the horizon. Thus, near the event horizon, 
the scalar field action becomes
\begin{equation}
 \begin{split}
  S[\varphi]=&\frac{1}{2}\int d^D x \sqrt{-g}\,\varphi \nabla^2 \varphi\\
=& \frac{1}{2}\int d^D x \,r_+^{D-4}\sqrt{\gamma}\,\mu(r_+^2+a^2)\\
&\times \varphi
  \left[-\frac{2M}{V-2M}\left(\partial_t -
  \frac{a}{r_+^2+a^2}\partial_\phi\right)^2 \right.\\ 
&\qquad\qquad\qquad\qquad\qquad +\left. \partial_r \frac{V-2M}{2M}\partial_r \right]\varphi  \ .
 \end{split}
\label{eq:My2dim}
\end{equation}
Let us make the following transformation
\begin{equation}
 \begin{cases}
 \psi = \phi + \frac{a}{r_+^2+a^2} t \ , \\
 \xi = t  \ . 
\end{cases}
\end{equation}
The result is given by
\begin{equation}
 \begin{split}
  S[\varphi]
=& \frac{(r_+^2+a^2)r_+^{D-4}}{2}\int d^D x\sqrt{\gamma}\mu\,\varphi(-\frac{1}{f}\partial_{\xi}^2
  + \partial_r f \partial_r)\varphi\\
=& \frac{(r_+^2+a^2)r_+^{D-4}}{2}\\
&\times\sum_n \int d\xi dr\,\varphi_n
  (-\frac{1}{f}\partial_{\xi}^2 + \partial_r f \partial_r)\varphi_n  \ ,
 \end{split}
\label{eq:My2dim}
\end{equation}
where
\begin{equation}
 f(r)=\frac{V-2M}{2M} \ .
\end{equation}
In the last line, we expanded $\varphi$ using the complete set of 
orthonormal functions of 
($\theta,\theta_2,\dots,\theta_{n+\epsilon-1}$,
$\psi,\phi_2,\dots,\phi_n$)
with the measure $\sqrt{\gamma}\mu$.
Eq. (\ref{eq:My2dim}) is the action for infinite set of scalar fields in 
the 2-dimensional spacetime with the metric 
\begin{equation}
 ds^2 = -f(r)d\xi^2 + \frac{1}{f(r)}dr^2  \ .
\end{equation}

Using the procedure of Sec.\ref{sec:level1}, we can get the correct Hawking
temperature of Myers-Perry black holes as
\begin{equation}
 T = \frac{V'(r_+)}{8\pi M} 
 = \frac{(D-3)r_+^2 + (D-5)a^2}{4\pi r_+(r_+^2+a^2)}  \ .
\end{equation}

\section{conclusion}
We have shown that the Hawking radiation from Kerr blacks holes can be
explained from gravitational anomalies point of view. The key
to show this is the dimensional reduction of the scalar field in the
Kerr spacetime. This result has been further generalized to 
Myers-Perry black holes with the single angular momentum.
We have also argued the validity of the boundary condition
by comparing the gravitational anomaly method with the trace anomaly method
in the case of 2-dimensional black holes. 
Thus, we have given an evidence of the universality of the gravitational
anomaly method.

It is interesting to extend the present method to other
 tensor fields such as the gravitons. 
 If we succeed to explain the Hawking radiation
 for these fields using gravitational anomaly method, 
 our understanding of black hole physics would become more profound. 

Some of recent researches on counting the black hole entropy 
are also related to anomalies 
\cite{Dabholkar:2004yr,Kraus:2005vz,Carlip:2006fm}.
It is interesting to give a unified view for both the black hole entropy
and  the Hawking radiation from gravitational anomalies point of view.
In particular, incorporating the back reaction of Hawking radiation
into the framework of the gravitational anomaly method is more challenging.

\vskip 0.5cm
Note: while this manuscript was prepared for submission, we noticed 
the independent work by S.Iso, H. Umetsu, and F.Wilczek \cite{Iso:2006}
on the same subject. A related work by Elias C. Vagenas and Saurya Das
\cite{Vagenas:2006qb} appeared on the
archive on the same day we submitted our paper.

\begin{acknowledgements}
This work is supported by the Grant-in-Aid for the 21st Century COE "Center for Diversity and Universality in Physics" from the Ministry of Education, Culture, Sports, Science and Technology (MEXT) of Japan. J.S. is supported by
the Japan-U.K. Research Cooperative Program, the Japan-France Research
Cooperative Program, the Grant-in-Aid for  Scientific
Research Fund of the Ministry of Education, Science and Culture of Japan 
 No.18540262 and No.17340075.  
\end{acknowledgements}

\appendix

\section{Comparison with Trace anomaly method}

We examine the validity of the boundary condition proposed by \cite{Iso:2006wa}
in the context of 2-dimensional Schwarzschild black hole. 
\begin{equation}
 ds^2 = -\left(1-\frac{2M}{r}\right)dt^2
  + \left(1-\frac{2M}{r}\right)^{-1}dr^2 \ .
\end{equation}
We will show that the gravitational anomaly method with the above
boundary condition gives the 
same result as the one derived using the trace anomaly
\cite{Christensen:1977jc,Mottola:2006ew}. Let us consider the massless scalar field theory
in the 2-dimensional Schwarzschild black hole
\begin{eqnarray}
  S = -\frac{1}{2} \int d^2 x \sqrt{-g} g^{\mu\nu}
             \partial_\mu \varphi \partial_\nu \varphi \ .
\end{eqnarray}
It is known that trace anomaly appear in this theory  as
\begin{equation}
 T^\mu_\mu = \frac{R}{24\pi} .
\end{equation}
From this trace anomaly, one can obtain the quantum effective action 
\begin{equation}
 W[g] = -\frac{1}{96\pi}\int d^2x \sqrt{-g}R\frac{1}{\Box}R \ .
\end{equation}
Using the auxiliary field $\chi$, we can rewrite this effective action 
as 
\begin{equation}
 W[\chi,g] = -\frac{1}{96\pi}\int d^2x \sqrt{-g}(-\chi \Box
  \chi + 2\chi R) \ .
\end{equation}
The energy momentum tensor is given by
\begin{equation}
\begin{split}
 \langle T_{\mu\nu} \rangle=&-\frac{2}{\sqrt{-g}}
 \frac{\delta W[\chi,g]}{\delta  g^{\mu\nu}}\\
 =& \frac{1}{48\pi}[\nabla_\mu \chi \nabla_\nu \chi - 2\nabla_\mu
 \nabla_\nu \chi \\
&+ g_{\mu\nu}\{2R - \frac{1}{2}(\nabla \chi)^2\}] \ .
\end{split} 
\end{equation}
The equation of motion for $\chi$ is
\begin{equation}
 -\left(1-\frac{2M}{r}\right)^{-1}\partial_t^2 \chi +
  \partial_r \left(1-\frac{2M}{r}\right)\partial_r \chi = \frac{4M}{r^3} \ .
\end{equation}
The solution of this equation is
\begin{equation}
 \chi = at - \log\left(1-\frac{2M}{r}\right) + A(r+2M\log(r-2M))+B
\end{equation}
where $a,A,B$ are constants. Using this $\chi$, the energy momentum tensor
is
\begin{equation}
 \begin{split}
  \langle T_{tt} \rangle =&
  \frac{1}{12\pi}\left(-\frac{2M}{r^3} + \frac{7M^2}{2r^4}\right) +
  \frac{1}{48\pi}\frac{A^2+a^2}{2}  \ ,\\
  \langle T_{rr} \rangle =&
  -\frac{1}{48\pi}\left(1-\frac{2M}{r}\right)^{-2}\left(\frac{2M^2}{r^4}
  - \frac{A^2+a^2}{2}\right)  \ ,\\
  \langle T_{rt} \rangle =& \frac{1}{48\pi}\frac{Aa}{1-\frac{2M}{r}} \ .
 \end{split}
\end{equation}
We adopt Unruh vacuum condition \cite{Unruh:1976db}, ingoing
flux should be vanish at $r=\infty$ and free fall
observer should observe regular energy momentum at $r=2M$. That is
\begin{equation}
\begin{cases}
 \langle T_{vv} \rangle = 0\qquad (\text{at } r=\infty)  \ ,\\
 \langle T_{uu} \rangle = 0\qquad (\text{at } r=2M)  \ ,
\end{cases}
\end{equation} 
where
\begin{equation}
\begin{cases}
 u=t-r-2M\log \left(\frac{r-2M}{2M}\right)  \ , \\
 v=t+r+2M\log \left(\frac{r-2M}{2M}\right)  \ .
\end{cases}
\end{equation}
From this boundary condition, we obtain $A=-a=1/4M$. 
Thus, we get the flux of Hawking radiation
\begin{equation}
 {T^r}_t = -\frac{1}{768\pi M^2}\equiv -\Phi  \ .
\end{equation}
Comparing this result with Eq. (\ref{eq:EM}), we get $K_o=\Phi$.
That is $K_H=0$ from Eq. (\ref{eq:constsrel}). This means that the boundary
condition proposed in \cite{Iso:2006wa} is justified at least
in the case of 2-dimensional Schwarzschild black hole. 
This supports the validity of the boundary condition 
$\tilde{T}^r{}_{t}=0$ at the horizon in any dimension.

\end{document}